\documentclass[journal,12pt,onecolumn]{IEEEtran}
\usepackage[top=2.0cm, bottom=2.2cm, left=2.2cm, right=2.2cm]{geometry}%Makes 2.5cm margins on my printer
\ifCLASSOPTIONonecolumn
\usepackage{setspace}
\fi

\input arraymac.tex
\usepackage{amsmath}
\usepackage{amssymb}
\usepackage{amsthm}
\usepackage{eqnarray}
\usepackage{graphicx}
\usepackage{wrapfig}
\usepackage{algorithmic}
\usepackage{cite}
\usepackage{comment}
\usepackage[font=small,labelfont=bf]{caption}
\usepackage{subcaption}
\usepackage{xcolor}%{}. Contents of {} are limited to no more than a paragraph. Many color names work.

\usepackage{etoolbox}
\makeatletter
\patchcmd{\@makecaption}
  {\scshape}
  {}
  {}
  {}
\makeatother
\def\tablename{Table}
\renewcommand{\thetable}{\arabic{table}} %Changes Table numbering from roman to arabic

\newcommand{\subparagraph}{}
\usepackage[compact]{titlesec}
\titlespacing{\section}{0pt}{0.9ex}{0.7ex}
\titlespacing{\subsection}{0pt}{0.7ex}{0.7ex}
\titlespacing{\subsubsection}{0pt}{0.7ex}{0.7ex}

\newcommand{\mycomment}[1]{%
}%
\usepackage{environ}
\NewEnviron{nestedcomment}{\mycomment{\BODY}}
\def\urltilde{\kern -.15em\lower .7ex\hbox{\~{}}\kern .04em}

\allowdisplaybreaks
\begin{document}
\title{Wave-Controlled Metasurface-Based Reconfigurable Intelligent Surfaces}
\author{\IEEEauthorblockN{\normalsize Ender~Ayanoglu,~\IEEEmembership{\normalsize Fellow,~IEEE,}
}\IEEEauthorblockN{Filippo~Capolino,~\IEEEmembership{\normalsize~Fellow,~IEEE,}
}~and~\IEEEauthorblockN{A.~Lee~Swindlehurst,~\IEEEmembership{\normalsize~Fellow,~IEEE}
}\\
\thanks{The authors are with the Center for Pervasive Communications and Computing, Department of Electrical Engineering and Computer Science, University of California, Irvine, CA 92697-2625, USA. This work is partially supported by NSF grant 2030029.}}
\normalsize
\maketitle
\begin{abstract}
Reconfigurable Intelligent Surfaces (RISs) are programmable metasurfaces that can adaptively steer received electromagnetic energy in desired directions by employing controllable phase shifting cells. Among other uses, an RIS can modify the propagation environment in order to provide wireless access to user locations that are not otherwise reachable by a base station. Alternatively, an RIS can steer the waves away from particular locations in space, to eliminate interference and allow for co-existence of the wireless network with other types of fixed wireless services (e.g., radars, unlicensed radio bands, etc.). The novel approach in this work is a wave-controlled architecture that properly accounts for the maximum possible change in the local reflection phase that can be achieved by adjacent RIS elements. It obviates the need for dense wiring and signal paths that would be required for individual control of every RIS element, and thus offers a substantial reduction in the required hardware. We specify this wave-controlled RIS architecture in detail and discuss signal processing and machine learning methods that exploit it in both point-to-point and multi-cell MIMO systems. Such implementations can lead to a dramatic improvement in next-generation wireless, radar, and navigation systems where RIS finds wide applications. They have the potential to improve the efficiency of spectrum utilization and coexistence by orders of magnitude.
\end{abstract}
\section{Introduction}\label{sec:backgroundmaterial}
\subsection{Reconfigurable Intelligent Surfaces}\label{sec:hmimos}
A Reconfigurable Intelligent Surface (RIS) is a surface used for enhancing wireless communication.
These surfaces may actively generate beamformed RF signals or control reflections of RF signals generated at other locations. Compared with standard RF transceivers, they are implemented with reduced cost, size, weight, and power consumption and transform the wireless environment into a smart, programmable entity. Because an RIS does not require downconversion of the received waveforms for baseband processing, there is no added thermal noise. RISs can be implemented from  low GHz frequencies up to the THz range where the miniaturization of electronics may provide integrated implementations.
RIS links have very low latency, while amplify-and-forward relaying can introduce delay.
Compared with standard RF transceivers, RISs are composed of inexpensive passive elements.

RIS systems can be planar structures with a thickness of a few centimeters, and they can be deployed in both outdoor and indoor scenarios, on building walls, stadium structures, in shopping malls, airports, etc. They can be employed to extend the range of the basestation (BS), connecting users with a blocked propagation path by reflecting electromagnetic waves from the BS around obstacles and in the direction of the user equipment (UE).
%%%, as in a relay station.
This can include extending BS coverage from outdoors to indoors.

Research on RISs began with the study of metasurfaces.
Metasurfaces have been proposed for a variety of applications, from high impedance surfaces
that perform as artificial magnetic conductors
to improve antenna efficiency,
as holographic leaky wave antennas,
as polarization convertors,
as reflect or transmit lenses,
as sensors,
or as analog devices that perform mathematical operations in real time.
Very recently research has focused on showing that tunable, controllable metasurfaces may improve communication links.

\begin{figure}[!t]
\centering
\ifCLASSOPTIONonecolumn
\includegraphics[width=0.55\textwidth]{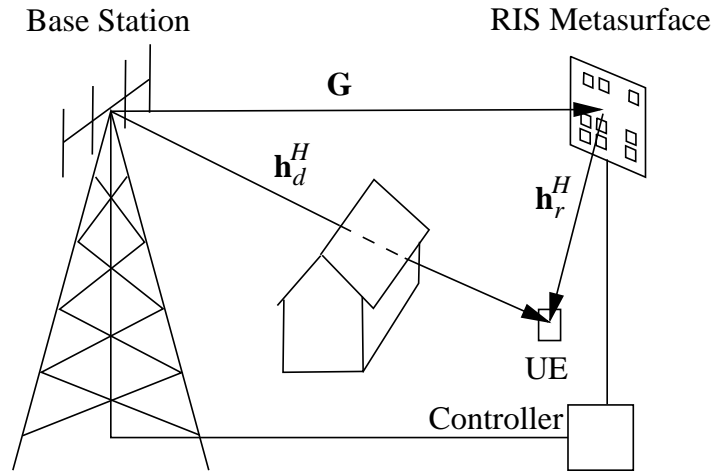}
\else
\includegraphics[width=0.35\textwidth]{Figures/channels}
\fi
\caption{RIS-aided wireless system.}
\label{fig:channels}
\end{figure}
In Fig.~\ref{fig:channels} we
show a BS employing an array of
antenna elements, and an RIS with $N_x$ elements in each row and $N_y$ elements in each column, for a total $N=N_xN_y$ elements. For now, we will assume that these elements introduce only phase shifts that can be individually controlled.
Each element of the RIS introduces a phase shift $\theta_n$
which we collect in the main diagonal of the diagonal matrix ${\bf\Theta}$.
Optimization algorithms have been formulated to maximize the signal-to-noise ratio (SNR) of the link subject to the physical properties of ${\bf \Theta}$, but such optimizations are non-convex and NP-hard. A number of attempts have been made to improve the computational complexity but their solutions deviate significantly from the optimum solution or creating a loss in SNR spectral efficiency. In Section~\ref{sec:ML4HMIMO} and \ref{sec:thrust2}, we will discuss potential methods to solve this problem.
\subsection{Tunable and Controlled Metasurfaces}\label{sec:wavemimo}
An RIS can establish favorable wireless channel responses by controlling the multipath and the diversity of the wireless propagation environment through its reconfigurable passive elements. % \cite{WZ18,WZ19}.
Prior work in the communication theory literature has assumed that the metasurface elements can be {\it individually and arbitrarily} controlled for this purpose. However, a critical observation is that each element of a metasurface is electromagnetically coupled to its neighboring elements, via the dielectric substrate and via free space, including the strong nearfield, and such coupling must be taken into account, together with other constraints and hardware limitations. Furthermore, most existing work on reflecting metasurfaces assumes that each individual metasurface element totally reflects the incident power with an idealized phase shift with respect to the incident wave, but in practice this is not possible due to electromagnetic coupling and losses.
Some previous studies such as \cite{R1} include analyses of the impact of hardware impairments, but such work still relies on idealized assumptions. RIS limitations should be carefully assessed using full-wave electromagnetic simulations of specific implementations \cite{Taghvaee2020}. Such full-wave simulations are particularly challenging due to the large dimensions of the RIS in terms of wavelengths and authors have also suggested hybrid methods inspired by previous studies on reflectarrays \cite{huang2007reflectarray}.\footnote{References \cite{R2} and \cite{R3} carry out an analysis on performance from the viewpoint of the impact of reflection loss.}

One of the simplest ways to realize tunable metasurfaces in the GHz range is through the use of varactor diodes that provide a tunable capacitance via a biasing voltage. This biasing voltage must be modulated over time (on the order of milliseconds) in order to change the reflectance properties of the metasurface, accommodating UEs with time-varying channel conditions. Tunability by means of varactor diode biasing has been proposed since the early days of research on metasurfaces \cite{SSSLT03}, and it is the easiest and most natural approach. Various realizations have been proposed connecting one or two varactors to each element of the metasurface. For example, arrayed metallic patch elements on a grounded dielectric slab can be connected to two varactors, each of which controls a reflected polarization. Different designs can lead to various ranges of phase shifts and relatively stable reflection coefficient amplitudes across frequency and phase.

Futuristic RIS implementations at millimeter wave and THz frequencies may involve other tuning mechanisms like graphene, liquid crystals, Schottky diodes, etc. The proposed approach using wave-controlled (large-domain) basis functions  is compatible with these technologies, and may even be more useful for RISs working at millimeter waves and THz because the control of each RIS element is even more challenging due to reduced dimensions and the high level of integration.

While prior work has emphasized exerting control on each element of the metasurface, we find this both unnecessary and expensive in terms of fabrication, implementation, and programmability. For example, a metasurface composed of $100\times 100$ elements would require signal paths reaching (via wires or printed circuits) each of the 10,000 unit cells. This doubles if we assume dual polarization operation. Furthermore, such control requires an optimization over 10,000+ parameters. The main contribution of this paper is to introduce an alternative solution that would substantially simplify this approach. This approach employs standing waves on the surface of RIS instead of 10,000+ individual wires or printed circuits. The resulting system is called wave-controlled metasurface-based RIS and will be discussed in detail in Section~\ref{sec:proposed}.
\subsection{Machine Learning for RIS}\label{sec:ML4HMIMO}
The majority of prior machine learning (ML) work on RISs has focused on their use in a single-cell scenario, i.e., a single BS with a single RIS serving one or more UEs. As cells become smaller and more dense, it is likely that an RIS, which could cover a large area on a building or signage, would be simultaneously visible to multiple BSs. In addition, UEs would also be simultaneously visible to multiple RISs.
As an RIS makes changes to the propagation environment, wireless channels seen by multiple BSs and UEs will also be affected. Thus, not only the channel of interest, but also the interfering channels will be impacted.
Coordination of multiple RISs and BSs in order to improve system coverage and capacity is a complicated problem with thousands of interrelated variables.
This could be addressed by simpler RISs, but it is worthwhile to consider ML approaches that
can provide solutions to intractable model-based approaches.

\begin{figure}[!t]
\centering
\ifCLASSOPTIONonecolumn
\includegraphics[width=0.65\textwidth]{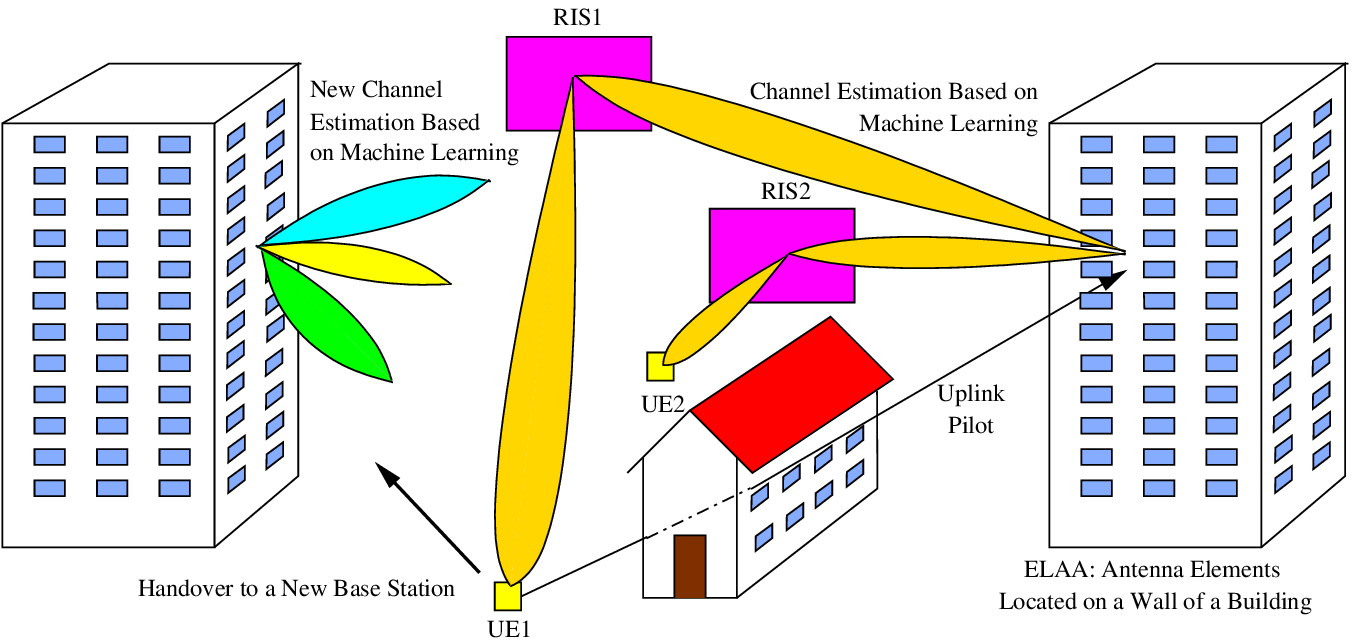}
\else
\includegraphics[width=0.45\textwidth]{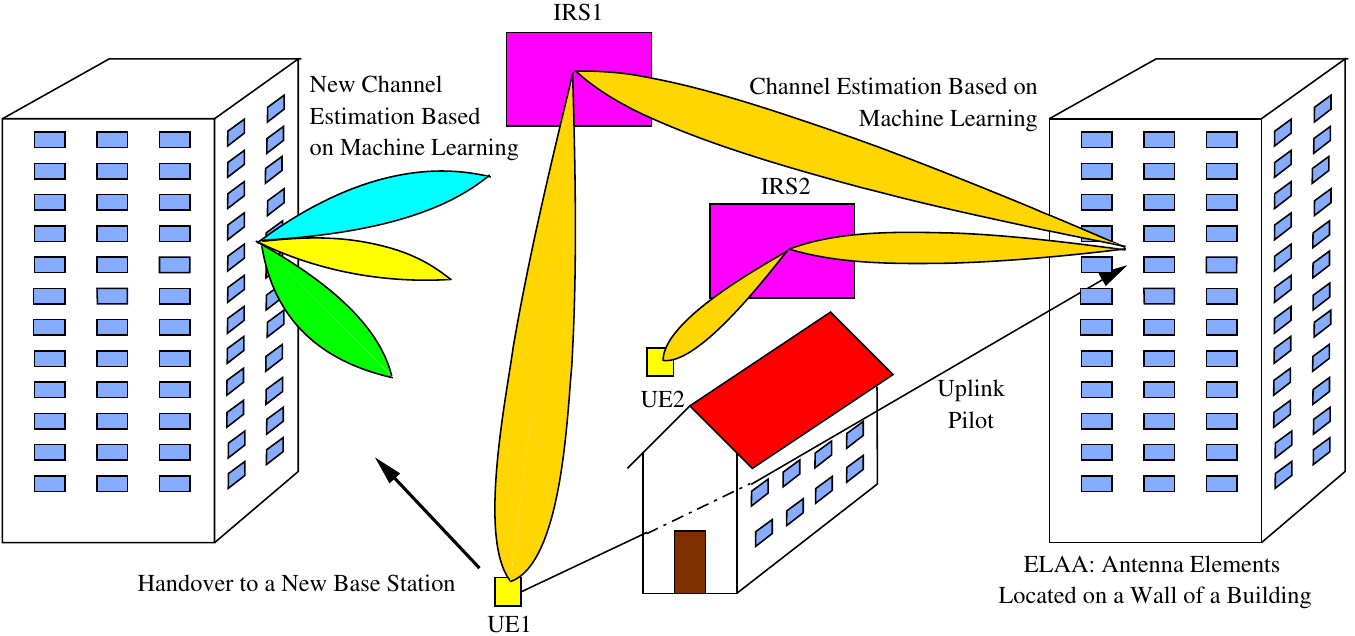}
\fi
\caption{Use of ML for RIS. The presence of the building forces the BS to employ RIS1 to reflect the downlink signal to reach the user when its LoS path is blocked. This redirection operation can be initiated based on UE1's location and tracked over a three-dimensional map in real time. In addition, at points along UE1's trajectory, BS handovers will be required, which can be determined by means of ML algorithms based on a combination of location, signal strength, and predictions of UE1's mobility.}
\label{fig:hmimo}
\end{figure}
ML techniques are known to be useful when a good model for a problem does not exist, the model cannot be mathematically analyzed, or algorithms based on the model are prohibitively complex. Many of the problems one can expect to address fit one or more of these criteria. ML techniques can be employed for the operation of RISs in a number of different areas, as depicted in Fig.~\ref{fig:hmimo}.

A basic requirement in all MIMO applications is calculation of the instantaneous channel state information (CSI). For implementations involving large numbers of antenna elements, the acquisition of this information becomes a big task. For such applications, acquisition and storage of CSI for large durations of time, together with side information such as user position, system characteristics via transmitter fingerprinting, %\cite{HC18,CGM20},
channel states over a number of different frequencies, etc., are very valuable for performance improvement in many applications. ML techniques are a valuable tool for extracting useful information from big data, and we expect that large systems employing a mixture of BSs, extremely large antenna arrays (ELAAs) and RISs will generate vast amounts of data that can be extracted to improve system performance. Decisions to employ RISs for particular UEs will involve the UE locations as well as their likely travel routes.
\section{Research Directions}\label{sec:proposed}
\subsection{Wave-Controlled RIS}\label{sec:thrust1}
\subsubsection {Estimation of Physically Realizable Cell-to-Cell Control of Reflection Properties}\label{sec:2.A.1}

\begin{figure}[!t]
\vspace{-1mm}
\centering
\ifCLASSOPTIONonecolumn
\includegraphics[width=0.65\textwidth]{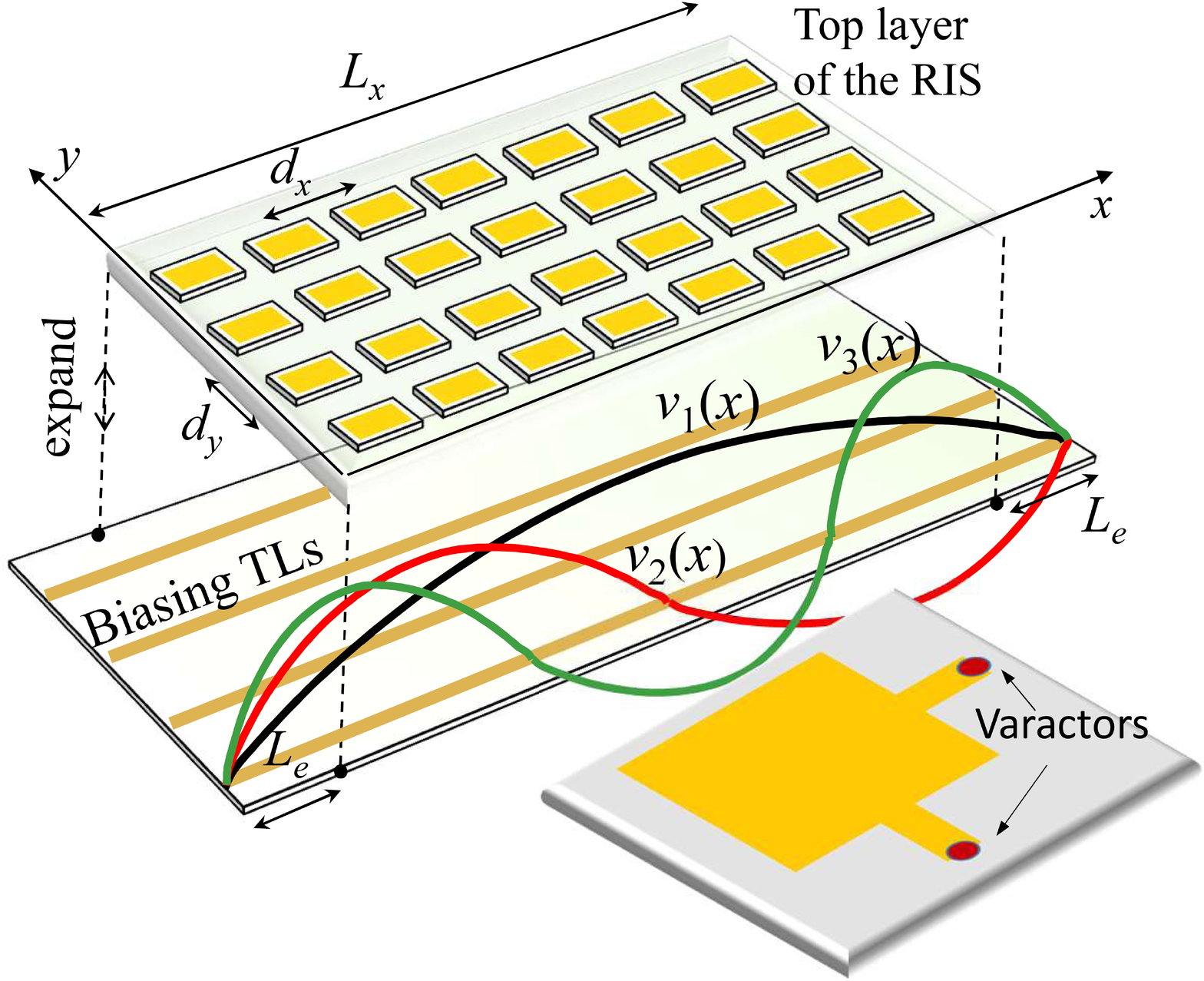}
\else
\includegraphics[width=0.45\textwidth]{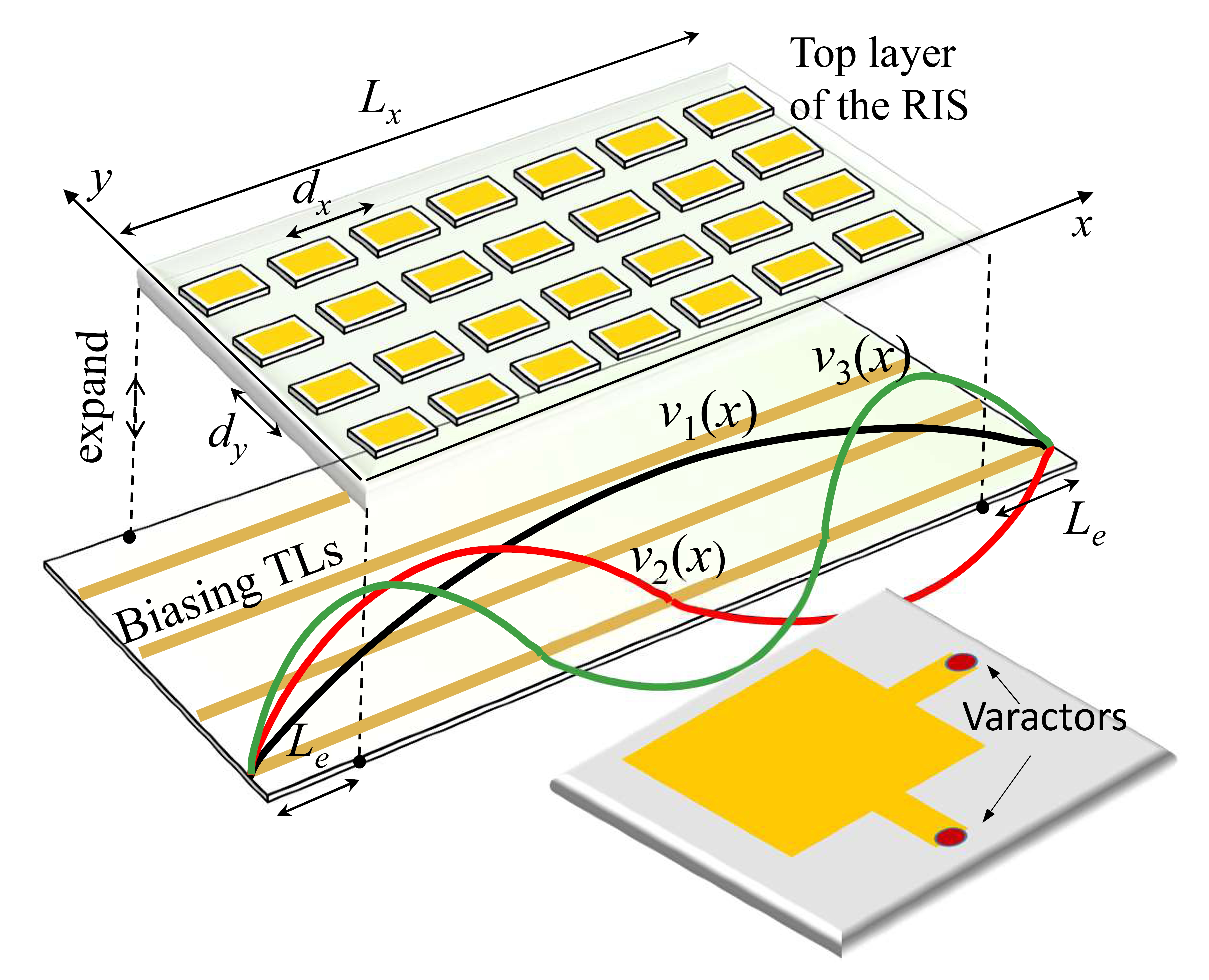}
\fi
\caption{A wave-controlled RIS provides simplified control of reflection properties. On the top layer the patches provide controlled wave reflection, at the bottom layers the biasing TLs provide the control. Instead of independently controlling each element of the metasurface, a reduced set of full-domain basis functions represented here as a set of voltage standing waves are used for biasing varactors. As an example, RIS elements made of  a dual-pol patch with varactors is shown; alternatively the varactor control can be exerted on the patch feeding lines on the bottom level (not shown).}
\label{fig:WaveControlledIRS}
\end{figure}

The reflection properties of the RIS have to be modulated in space across the reflecting surface.
The main underlying concept is the variation of the RIS's local surface impedance and its transverse gradient; in addition to using electric voltage, tuning can be done through other means, including thermal excitation, optical pumping, etc. In the low GHz range,
the properties of the unit cells are %%%to be
varied across the surface by tuning varactors, acting mainly as variable capacitors, that are connected to each unit cell metallic element, like a patch antenna shown in Fig. \ref{fig:WaveControlledIRS}.

As mentioned in Section~\ref{sec:wavemimo}, it is not possible to control the electromagnetic-field reflection properties from cell to cell of a metasurface, even when a varactor is connected to every unit cell. Indeed it is not even proper to discuss the problem in terms of individual-cell ``reflection'' because reflection is a collective phenomenon generated by the constructive interference of radiation arising from several unit cells.
The concept of the ``unit cell reflection property'' is the reflection coefficient evaluated by a fully periodic structure, i.e., with all unit cells equivalent to each other (i.e., a phase gradient). This term is also used when a metasurface is generating a change of direction in the reflected beam, according to the generalized Fermat principle or generalized Snell law.
Control of the reflected beam direction is obtained by applying a proper linear phase shift across the metasurface dimension. For example, when a beam is incident from
a given direction, the reflected beam can propagate in a different direction.

Here we do not simply aim at generating a beam deflection, we aim at having a metasurface that reconfigures itself, and its scattering/reflection properties are much more complicated than those previously studied, i.e., they can generate multiple beams or even complex radiation patterns to equalize the channel, including polarization control.
When different voltage biases are applied to varactors in adjacent unit cells (Fig.~\ref{fig:WaveControlledIRS}), a desired phase shift of the ``locally'' reflected field is produced, but the actual phase shifts across the metasurface cells do not follow the same arbitrary voltage variation that is applied to the varactors. In other words, the realized phase shift of the locally reflected fields across the surface is smoothed out because of all possible couplings via free space and via the dielectric substrate. These couplings usually occur in the so-called ``nearfield'' and therefore can be very significant.

Metasurfaces usually are very large in terms of wavelengths, involving many small-scale (subwavelength) features, and therefore numerical simulations are very time consuming. A typical approach has been to simulate a single-periodic unit cell. This approach is very accurate when the metasurface is ``uniform'' in its excitation, i.e., when the excitation across the surface is linearly phased (also called pseudo periodic excitation, like that provided by a plane wave). However, in many applications, various parts of the metasurface must provide different reflection characteristics and therefore, properly speaking, a numerical method using periodic boundary conditions is not necessarily accurate. On the other hand, the scheme we will describe is based on having a reduced number of degrees of freedom that provide a smoother phase function
across the surface. Indeed, we aim at modulating (in space) the reflection properties not from cell-to-cell but rather over a larger scale, from groups-to-groups of cells. Therefore the so called ``local periodicity condition'' used in the numerical full-wave modeling of the metasurface, already widely used in the design of reflectarrays \cite{ZSJMKB11}
and metasurfaces \cite{MCMSM16}, is expected to be accurate.
\subsubsection {Robust but Simplified Control of Reflection Properties}\label{sec:2.A.2}
Once the number of degrees of freedom to generate the necessary RIS field scattering is determined for a few representative metasurface geometries, one can focus on modeling long metasurfaces in one direction, say $x$, while in the other dimension, say $y$, the metasurface is assumed to be infinitely long so a single unit cell in the $y$ direction is numerically simulated using full-wave methods. This methodology enables the study of the effects of the locally reflected-field variation in a much faster way than in a study of rectangular metasurfaces, with finite-lengths in both directions, without  compromising the generality of the results. We propose to apply a set of ``full-domain'' basis functions to control the phase shift of each unit cell across the RIS in the $x$ direction. For example, if the RIS has $N_x$ unit cells in the $x$ direction, a maximum number $N_x$ of individual controls could be
applied. However, we argue that when using full domain basis functions a significantly smaller number of degrees of freedoms is needed to implement the proposed functionality.
The rationale is that {\em i)\/} the phase shift from cell to cell of the RIS is smoothed out in the Fresnel zone and even more in the farfield (Fraunhofer) region due to the low-wavenumber-pass filtering and wavenumber asymptotic localization caused by electromagnetic propagation; {\em ii)\/} as mentioned in Sec.~\ref{sec:2.A.1}, the nearfield couplings from cell to cell limit the scanning angle of operation anyway (a phenomenon well known in large scanning angle arrays \cite{mailloux2018phased}), and hence attempts at implementing a true cell-to-cell control would not be meaningful. This method has the potential to greatly simplify the design of RIS and because of its possible high impact, the accurate estimate of degrees of freedom reduction should be the subject of future investigations.

\subsubsection {Design and Study of Scattering Freedom Using a Programmable RIS}\label{sec:II-A-3}
Various possible designs that incorporate the wave-controlled mechanism described above are possible. For example, the design
can consist of a laminated metasurface where the top layer has a periodic arrangement of unit cells made of patterned copper patches attached to a varactor, which is connected to a biasing line below the ground plane as shown in Fig.~\ref{fig:WaveControlledIRS}. The bias is administered using transmission lines (TLs) that do not interfere with the incoming and reflected RF waves. The TL used for biasing the varactors (hereafter referred to as the ``biasing TL'') supports (biasing-) modes that travel with phase velocity $\nu_b$ over the whole length $L_x$ of the metasurface along the $x$ direction. A biasing mode has a TL-propagation wavenumber $k_b$
and guided wavelength $\lambda_b$.
One can use a number $P_x$ of biasing modes
where $P_x$ is the total number of degrees of freedom discussed previously used to control the scattering properties of the RIS.

An important issue to consider is
the power consumption of biasing lines closed on the loads and also the option of closing the biasing TL on short circuits, and how these two solutions affect the flexibility in assigning proper biasing to the varactors. When the biasing TL is closed on short circuits, a biasing-standing wave is established. To avoid short-circuiting the first and last cells of the metasurface, the biasing line may be chosen to be longer than $L_x$ (the size of the RIS) by a small extra length $L_e$ on each side, as shown in Fig.~\ref{fig:WaveControlledIRS}.
The first three standing waves
that enable the wave-control are shown in Fig.~\ref{fig:WaveControlledIRS}. The proper frequency of excitation $f_p$
of the standing wave will be determined by establishing a resonance that, neglecting loading along the TL, satisfies the
boundary conditions.
The goal is to use an RIS with size  $L_x^{\prime}$ much larger than a wavelength, so that the resonance frequencies are much smaller than the electromagnetic RF frequency used in the communications link; this simplicity enables the wave-controlled method proposed here, and hence the use of the reduced degrees of freedom.

We assume that the standing wave is used to provide a polarization bias to the varactors, located at positions $m d_x$. The time-domain field representation of the standing biasing wave is
\begin{equation}
v(x,t) =  V_0 +\sum_{p=1}^{P_x} V_p \sin(k_{b,p} x+\phi_{e,p}) \cos (\omega_p t +\phi_{v,p}),
\label{biaswave}
\end{equation}
which consists of the summation of $V_0$ and $P_x$ terms with the product of a sine function in the spatial domain and a cosine function in the time domain. The amplitude of each term is $V_p$ where $p$ varies from 1 to $P_x$. The wavenumber for the $p$-th sine function is $k_{b,p}$ and its spatial phase is $\phi_{e,p}$. The radian frequency for the $p$-th cosine function is $\omega_p$ and its temporal phase is $\phi_{v,p}$.
where we have added the constant DC bias voltage $V_0$ to polarize the varactors. This provides an envelope polarization voltage at each RIS element. By properly choosing the excitation of each biasing TL mode $V_p$, we can provide a large degree of variation to the varactor biasing after a proper detection method is used.

\subsection{Reduced Dimension Design for RIS}\label{sec:qmimo}
The wave-controlled RIS architecture discussed above has important implications on how the wireless system performance can be optimized; in fact, it is symbiotic with the types of propagation environments for which RISs have been proposed, as discussed below. One of the challenges associated with a system employing an RIS is to determine the best values for the phase shifts of the RIS cells, and then communicating these values to the RIS. Approaches that have been proposed to date attempt to determine each of the individual phase shifts independently of the others, assuming either that the phases are continuously adjustable, or are quantized to a resolution of a few bits. In either case, it has been shown that a large asymptotic SNR gain can be achieved, but finding the elements of $\Thetabf$ is typically posed as a large non-convex optimization problem that must be solved for each user and each channel coherence interval, and then transmitted to the RIS.

As argued above in Section~\ref{sec:thrust1}, achieving an arbitrary local phase shift at each cell of the RIS is not possible due to inherent electromagnetic coupling effects. We further argue here that such detailed control is in fact unnecessary,
especially in the types of situations for which RIS architectures have been typically proposed, where the RF environment is dominated by relatively few propagation paths (e.g., millimeter wave or THz frequencies).
As explained below, the structure inherent in the problem can provide a much simpler RIS operation.
\subsubsection{Reduced Dimension RIS Control}\label{sec:2.B.1}
\begin{figure}[!t]
\centering
\ifCLASSOPTIONonecolumn
\includegraphics[width=0.65\textwidth]{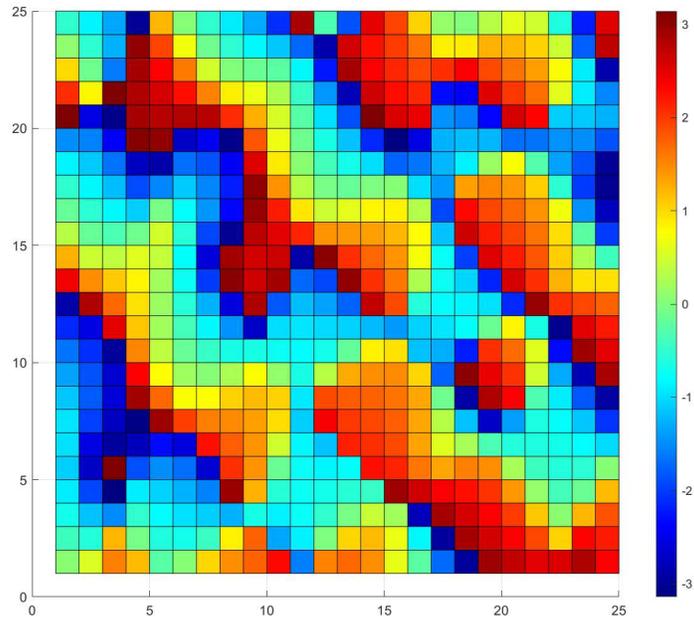}
\else
\includegraphics[width=0.45\textwidth]{Figures/phaseprofile3.eps}
\fi
\caption{Example of the optimal RIS phase at different cell positions.}
\label{fig:irs2d}
\vspace{2mm}
\end{figure}

Fig.~\ref{fig:irs2d} shows the optimal phase shifts for a case with $d_x=d_y=\lambda/5$, and a single-antenna transmitter and receiver communicating via the RIS. The channels from the transmitter to the RIS and from the RIS to the receiver are assumed to each be composed of 5 equal gain but random phase propagation paths with random angles of arrival/departure. It is clear that the change in phase from one RIS element to the next is relatively smooth, and hence that achieving an arbitrary variation in the phase across the RIS is unnecessary. Consequently, the required RIS phase profile can be described by many fewer than $N$ parameters. The above holds true also for cases involving large RISs with users located in the Fresnel region, as this does not fundamentally alter the smooth nature of the RIS phase from element to element, since the Fresnel region of the whole antenna can still be in the farfield zone associated with just a few RIS elements.\footnote{A discussion of farfield vs. nearfield operating properties of RISs can be found in \cite{BB}.}
Furthermore, given the arbitrary phase offset, the actual value of the phase of each element is not important; it is instead the phase difference between elements that determines the directionality of the resulting propagation. This smooth phase profile is fortuitous given the constraints described in Section~\ref{sec:thrust1} on the element-to-element phase variation due to electromagnetic coupling. Indeed, as explained in points {\em i)\/} and {\em ii)\/} in Sec.~\ref{sec:2.A.2}, the ideal control of the phase difference from cell to cell is not feasible due to physical constraints imposed by electromagnetic propagation and coupling. The limited number of biasing modes $P$ in our proposed control signaling is able to not only account for the maximum phase change that is physically realizable from cell to cell, it also reduces the dimension of the required optimization and the resulting information that must be communicated to the RIS. This is a critical issue since a separate optimization and control signal is required for every user with which the BS is using the RIS to communicate.

Unlike prior work that proposes direct design of the RIS element phases, in the proposed architecture it is the bias voltage excitations that are designed to achieve the desired response, which takes into account the element-wise coupling. A smooth phase profile composed of a sum of a small number of two-dimensional sinusoids can in principle be achieved by an intelligent superposition of the modes of the biasing control in~\eqref{biaswave}. The asymptotic gain in SNR achievable by the proposed architecture remains to be determined, including the case where the bias voltages are quantized.
Another key practical factor to consider is how to account for the boundary conditions on the bias voltages that will in turn limit the choice of the possible basis functions. As mentioned in %Thrust~1,
Section~\ref{sec:thrust1},
a possible solution is to increase the length of the waveguides beyond the outermost cell edges, to force the boundary conditions to occur in positions where they do not significantly constrain the bias voltage that can be achieved at the edge cells of the metasurface.

\subsubsection{Sparse UE Channel Parameter Estimation}\label{chanest}

It is well known that, in the absence of active transceivers at the RIS, one can only uniquely estimate the {\em composite} or {\em cascaded} channel of the overall BS-RIS-UE link, rather than the channels to and from the RIS. Methods that attempt to estimate the individual elements of the cascaded channel incur a large training overhead, requiring on the order of $N$ pilot symbols, which can be unacceptably large in many applications. In addition, such methods assume the ability to arbitrarily set the phase of the RIS elements, for example using DFT or Hadamard matrices, which is not realistic given the inherent electromagnetic coupling described above.

More recently, geometric channel models described by physical propagation parameters such as angles of arrival/departure and complex path gains have gained increased attention, due to the fact that the number of parameters required to determine the channel, and hence the amount of training overhead required, is typically much smaller than in the case of unstructured models. RISs are ideally suited for overcoming blockages in millimeter wave and THz channels which have sparse propagation paths, so low-dimensional geometric channel models are particularly relevant for the RIS application, and this is symbiotic with the idea of using low-dimensional biasing modes for RIS control.

Whereas the optimal design of the RIS element phases for training and beamforming has been addressed in prior work, a challenging open problem for the approach presented herein is how to best choose the biasing mode excitations in order to optimize the channel estimation and link performance. In both cases, the non-linear relationship between the biasing voltages and the RIS phase shifts leads to a difficult non-convex optimization problem. Fortunately, the required number of biasing modes will be relatively small -- although determining the precise number of such modes is also an open problem -- and though non-convex, the number of excitation parameters to estimate will be much smaller than $N$.
\subsection{CSI Acquisition and Radio Geometry Dimensionality Reduction via ML}\label{sec:thrust2}
For this system, it is possible to employ ML to attack two problems. The first is the determination of CSI and decision making as to BS handover based on CSI, while the second focuses on determination of the phase shift matrix $\Thetabf$. Going beyond the methods
described in Section~\ref{sec:qmimo}, it is possible to exploit the reduced-dimension wave-controlled model in developing algorithms that are targeted at situations with multiple {\em mobile} UEs, in which updates for the CSI and $\Thetabf$ are required within the channel coherence time. This requires considerations of higher-level issues not considered in Section~\ref{sec:qmimo}, but also offers some simplifications that can be exploited.

ML solutions for the control of RIS systems need to be considered for two reasons. First, with an RIS, the problem complexity increases substantially. And, second, ML techniques have been growing in power and are being used in many fields, including wireless communications.
Space limitations prevent us from a detailed discussion of potential ML techniques for RISs and their comparisons but \cite{DMPPQSYY20,LLMHXDA21} provide a good summary. Reference \cite{DMPPQSYY20} discusses {\em i)\/} using ML to assist CSI acquisition,  {\em ii)\/} using ML without explicitly calculating CSI, {\em iii)\/} combining model-based and ML approaches. Motivations for integrating ML in RISs are discussed in \cite[Section V.A]{LLMHXDA21}. Then, two ML techniques for RISs are discussed in detail: {\em i)\/} deep learning (DL) \cite[Section V.B]{LLMHXDA21} and {\em ii)\/} reinforcement learning (RL) \cite[Section V.C]{LLMHXDA21}. A very detailed comparison of different RL techniques from the literature is available in \cite[Fig. 5]{LLMHXDA21}. In \cite[Section V.E]{LLMHXDA21}, further discussion of other ML techniques that belong to the classes of {\em i)\/} supervised, {\em ii)\/} unsupervised, and {\em iii)\/} federated learning is provided. We believe it is fair to say that there is no clear ``winner'' ML technique for RISs. On the other hand, there is general agreement that it is needed. In passing, we emphasize that having multiple RISs in the environment and the complexity of their cooperation complicates the problem further. For example, the channel estimation problem becomes one of compound channel estimation.\footnote{This subject was discussed in Sec.~\ref{chanest}. A detailed treatment of it for THz channels is available in \cite{AA}.} On the other hand, we would like to emphasize that since the wave-controlled RIS described in this paper has substantially fewer parameters to control, its control through ML will be much simpler than conventional ML control of RISs.

\begin{figure}[!t]
\centering
\ifCLASSOPTIONonecolumn
\includegraphics[width=0.65\textwidth]{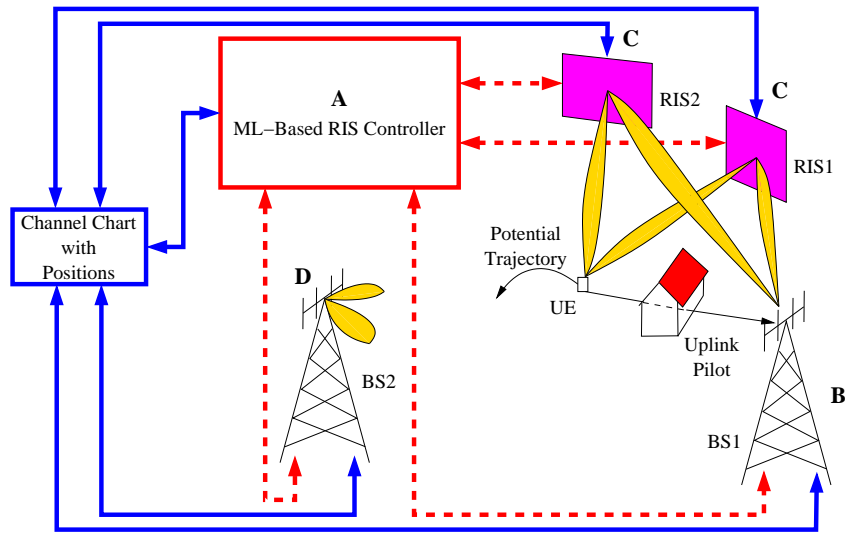}
\vspace{2mm}
\else
\includegraphics[width=0.45\textwidth]{Figures/MLRIS}
\fi
\caption{Opportunities for using ML in a network with RISs, control channels for BSs and RISs, and the use of channel charting %\cite{BSWHM19,DiRenzoetal19}.
\cite{DiRenzoetal19}. Channel charting can be used for UE localization, handovers, and for determination of CSI. Alternatively, other ML algorithms can be employed for a number of these tasks as described in the text. ML algorithms can be employed in A: control channel operation of the RISs, B: channel estimation, C: configuring RISs to reflect the signal towards the UE, and D: predicting trajectory changes and for optimized beam alignment by the target BS. It is important to realize that there will be much fewer control channels and therefore the task of managing the control channels will be much simpler in the wave-controlled metasurface-based RIS as compared to conventional RISs.}
\label{fig:MLRIS}
\end{figure}
It is possible to focus on a particular ML technique that employs extensive CSI but reduces its dimensionality to come up with what is termed a {\em channel chart\/} for a particular BS and the cell that it serves
\cite{SMGGT18}. A channel chart is a representation of the local ``radio geometry'' based on CSI in a substantially reduced dimensionality for a wireless cell. Channel charting is an unsupervised ML approach where learning is only based on CSI passively collected at a single point in space, from multiple transmit locations, over time
as in Fig.~\ref{fig:MLRIS}.
The method then extracts channel features that characterize large-scale fading properties of the wireless channel. In the next step, channel charts are generated.
Relationships related to the position and the movement of a UE in the cell can then be deduced from comparing measured radio channel characteristics to the channel chart. %, as shown on the bottom of Fig.~\ref{fig:channelchart3}.
\section{Conclusion}
In this paper, a new way of implementing RISs based on metasurfaces is introduced. The novelty of the approach is the fact that the phase shifts are governed not by individual controls that would require
numerous bulky electrical connections, but by electromagnetic waves launched onto thin waveguides on the reflecting surface. Given the boundary conditions of these waveguides, the controlling wave is designed as a linear superposition of periodic modes whose coefficients determine the phase shifts. This approach results in a reduced number of degrees of freedom that provide a smoother phase function across the surface. Implementation of the RIS, a reduced dimension design for it, and its operation by using ML techniques are discussed. The reduced degrees of freedom approach for biasing the elements that control the RIS phases will lead to lower cost and power
consumption. Optimizing only the biases with a fewer number of degrees of freedom than the phases will decrease the number of optimization variables and consequently reduce the complexity. The use of ML for channel charting and RIS phase optimization will provide an efficient solution for multi-cell multi-RIS approach. This RIS system can enhance the co-existence of various wireless systems through interference management and has the potential to improve the efficiency of spectrum utilization and coexistence by orders of magnitude.
\bibliographystyle{IEEEtran}
\bibliography{References/References,References/SpecEES,References/Metasurfaces_biblio}
\small
\section*{Biographies}
ENDER AYANOGLU [F] (ayanoglu@uci.edu) received his Ph.D. from Stanford University. He has been employed by Bell Laboratories Research and Cisco Systems. Since 2002, he has been a professor at the University of California Irvine, where he has served as the CPCC director and held its Endowed Chair. He has received two best paper awards from IEEE ComSoc. He has served as Chair of the ComSoc Communication Theory Committee, Editor-in-Chief of IEEE Transactions on Communications, and Founding Editor-in-Chief of IEEE Transactions on Green Communications and Networking.\\

FILIPPO CAPOLINO [F] (f.capolino@uci.edu) received his Laurea (cum laude) and Ph.D. degrees in electrical engineering from the University of Florence, Italy, in 1993 and 1997, respectively. He is currently a professor in the Department of Electrical Engineering and Computer Science of the University of California Irvine. His research interests include millimeter-wave antennas, metamaterials and their applications, traveling wave tubes, sensors in both microwave and optical ranges, wireless systems, chip-integrated antennas, and applied electromagnetics in general.\\

A. LEE SWINDLEHURST [F] (swindle@uci.edu) received his Ph.D. from Stanford University. He served as the associate dean for research at the University of California Irvine, and was formerly Vice President of Research at ArrayComm LLC. He is the founding Editor-in-Chief of the IEEE Journal of Selected Topics in Signal Processing and recipient of several major IEEE paper awards. His research interests are focused on multichannel signal processing for wireless communications, radar, and biomedical applications.
\end{document}